\documentclass[prd,superscriptaddress,showpacs,twocolumn,floatfix]{revtex4}
\usepackage{graphicx}
\usepackage{amsmath}
  \def\e{{\rm e}}
  \def\l{\left(}
  \def\r{\right)}
\begin{document}

\title{Estimate of the correlation signal between cosmic rays and 
BL Lacs in future data}

\author{D.S.~Gorbunov}

\affiliation{Institute for Nuclear Research of the Russian Academy of
Sciences, 60th October Anniversary prospect 7a, 117312 Moscow, Russia}

\author{P.G.~Tinyakov}

\affiliation{Institute for Nuclear Research of the Russian Academy of
Sciences, 60th October Anniversary prospect 7a, 117312 Moscow, Russia}

\affiliation{Service de Physique
Th\'eorique, Universit\'e Libre de Bruxelles, CP225, blv.~du Triomphe, B-1050
Bruxelles, Belgium}

\author{I.I.~Tkachev}

\affiliation{Institute for Nuclear Research of the Russian Academy of
Sciences, 60th October Anniversary prospect 7a, 117312 Moscow, Russia}

\affiliation{CERN Theory Division, CH-1211 Geneva 23, Switzerland}

\author{S.V.~Troitsky}

\affiliation{Institute for Nuclear Research of the Russian Academy of
Sciences, 60th October Anniversary prospect 7a, 117312 Moscow, Russia}

\begin{abstract}
The existing correlation between BL Lacertae objects (BL Lacs) and 
cosmic-ray events observed by HiRes experiment provide sufficient
information to formulate quantitatively the hypothesis about the flux
of neutral cosmic-ray particles 
originated from BL Lacs. We determine the potential of
future cosmic ray experiments to test this hypothesis by
predicting the number of coincidences between arrival directions  
of cosmic rays and positions of BL Lacs on the celestial sphere,  
which should be observed in the future datasets. We find that 
the early Pierre Auger data will not have enough
events to address this question. On the contrary, the final 
Pierre Auger data and the early Telescope Array data will be
sufficient to fully test this hypothesis. If confirmed, it would imply
the existence of highest-energy neutral particles coming from
cosmological distances.
\end{abstract}


\maketitle

%%%%%%%%%%%%%%%%%%%%%%%%%%%%%%%%%%%%%%%%%%%%%%%%%%%%%%%%%%%%%%%%%%%%%%%

\section{Introduction}

The origin of particles with energies exceeding $10^{19}$~eV
(hereafter ultra-high-energy cosmic rays, or UHECRs) is one of the
most intriguing questions of astroparticle physics (for a review, see
e.g. Ref.~\cite{reviews}). Within conventional-physics explanations
the existence of these particles is attributed to acceleration in yet
unknown astrophysical sites. Various candidates have been proposed
(see e.g. Ref.~\cite{Ancho:rev}, for a recent review). Poor (in
astronomical standards) angular resolution of the cosmic-ray
experiments makes it impossible to identify actual sources
directly. Instead, one has to rely on statistical methods.

A clean and indisputable way to perform a statistical analysis
is to formulate a hypothesis about the sources before the data are 
released. This was not possible to do with the first data sets. At
present, however, there exists enough data to specify the class of
likely sources. More important, new high-quality data are expected
from cosmic-ray experiments in the very near future. Before these data
are released, a hypothesis to be tested should be formulated and the
procedure to verify or falsify it should be defined. In particular, it
is important to define quantitatively which results of a particular
experiment would test the hypothesis at a given confidence
level. In this note we determine such a quantitative procedure. 

During the last ten years, many classes of astrophysical objects were
tested for positional correlations with arrival directions of UHECRs
(see Ref.~\cite{Comparative} for a comparative study). Significant
correlations were found~\cite{BL1,BL:GMF,BL:EGRET} with different
samples of BL Lacs from the Veron catalog \cite{Veron2001}. The
correlations with BL Lacs appear in new independent data sets as
well.  Most recently the correlations between bright confirmed BL Lacs
and the arrival directions of events observed by the HiRes stereo
experiment, were found \cite{BL:HiRes} and confirmed by an independent
method \cite{HiRes:BL}. This case is particularly clean for it
involves minimum of assumptions concerning the BL Lac sample and
propagation of UHECR particles (the latter are assumed neutral and 
propagate along straight lines). Thus, it is most suitable for a
rigorous test with the future data. The purpose of this paper is to
formulate the testing procedure before the data are released.

We specify the hypothesis in Sec.~\ref{sec:hypo} and formulate the 
procedure to test it in the following sections. In
Sec.~\ref{sect:eta}, we describe how to estimate the
experiment-independent quantitative parameter involved in the
hypothesis on the basis of a single experiment (HiRes stereo in our
case). In Sec.~\ref{sec:Predictions}, we estimate the sensitivity of
future experiments to the suggested hypothesis and determine which
results would allow either to exclude or to confirm it at a given
confidence level. We calculate the corresponding parameters for the
final AGASA, new HiRes stereo, Pierre Auger and Telescope Array data
sets.

%%%%%%%%%%%%%%%%%%%%%%%%%%%%%%%%%%%%%%%%%%%%%%%%%%%%%%%%%%%%%%%%%%%

\section{The hypothesis to be tested}
\label{sec:hypo}

The most precisely determined published arrival directions 
were provided by HiRes in the stereo mode~\cite{HiRes}. 
They recorded 271 events with energies
$E>10^{19}$~eV, which were found~\cite{BL:HiRes} to correlate with
bright (visual magnitude $m<18$) confirmed BL Lacs from the 10th
edition of the V\'eron catalog~\cite{Veron2001} --- the sample
previously identified~\cite{BL:GMF} as correlated with cosmic rays
detected by the AGASA and Yakutsk experiments. With the angular
resolution (in the stereo mode) of $0.6^\circ$, the strongest
correlation signal was expected at $0.8^\circ$ as determined by
Monte-Carlo simulations in Ref.~\cite{BL:HiRes}.  There were 11 pairs
``BL Lac -- cosmic ray'' found with the separation less than
$0.8^\circ$ while in average $3.5$ pairs were expected for a randomly
distributed sample. The angular separations observed are much smaller
than a typical deflection of a charged particle of that energy in the
Galactic magnetic field, so the correlation may only be explained by
the presence of a fraction of neutral primary
particles~\cite{BL:HiRes}.

One can summarize these results in the form of a hypothesis suitable for
testing by other experiments. The hypothesis consists of
three assumptions:
\begin{enumerate}
\item[i)] Some fraction of cosmic-ray events with energy
  $E>10^{19}$~eV corresponds to neutral primary particles. 
\item[ii)] A part of these neutral particles, or their progenitors,
are emitted by the confirmed BL Lacs with magnitude $m<18$ marked
``BL'' in the catalog~\cite{Veron2001}. There are 156 such
objects. The neutral particles from BL Lacs compose the fraction $\eta$ 
of the observed UHECR flux. 
\item[iii)]
  The observed cosmic-ray fluxes of these 156 sources are roughly equal.
\end{enumerate}

Several remarks concerning the assumptions i)--iii) are in order.
Firstly, we determine the value of $\eta$ from the HiRes data. 
Secondly, the energy threshold of $E=10^{19}$~eV adopted in i) was used
in Ref.~\cite{BL:HiRes} because only the arrival directions of the
events with $E\geq 10^{19}$~eV were published in
Ref.~\cite{HiRes}. Further studies~\cite{HiRes:BL} demonstrated the
presence of the correlations in an independent (unpublished) data set
with $E<10^{19}$~eV. This represents however a different claim which
we do not discuss here; it can be analyzed in a similar way. When
considering other experiments in what follows we assume the same cut
on energy. This assumption may be important for the validity of the
predictions, as the fraction of neutral particles (which exhibit the
neutral correlations) may vary with energy. If these variations are
strong, the energy dependence of the acceptance of the experiments has
to be taken into account even at the same energy cut. Also, it is
necessary that absolute energy calibrations of different experiments
match each other. Otherwise this should be taken into account when
selecting events for testing the hypothesis (it is worth noting that,
currently, the spectra obtained by different experiments do not
match).

Thirdly, the assumption ii) does not exclude other possible sources
of UHECRs and means only that BL Lacs emit {\em some} of the
cosmic-ray particles. We will see shortly that according to HiRes data
the fraction $\eta$ of neutral UHECRs attributed to BL Lacs is
actually small, of order 2\%.

Finally, the assumption (iii) is a technical one; it may be replaced
by any other particular conjecture on the relative cosmic-ray
brightness of BL Lacs in the sample. In fact, our analysis requires
only that in the flux of cosmic rays observed by a given
experiment the fraction of events which are associated with BL Lacs
(this fraction equals $\eta$ if the experiment sees the whole sky with
the uniform acceptance) is proportional 
to the number of BL Lacs covered by the acceptance area.

Our further logic is as follows.  From each cosmic-ray experiment, one
can put bounds on the fraction $\eta$ under the assumptions 
(i)--(iii). Two experiments could, in principle, put the bounds which
are, with some probability $p$, {\em inconsistent} with each
other. Leaving aside the possibility that one of the experiments is
wrong, this would mean the rejection of the set of assumptions at the
confidence level $p$.

\section{Estimating $\eta$ from the HiRes data}
\label{sect:eta}

Consider a given catalog of $M$ astrophysical objects (candidate
sources) with celestial coordinates $\{\alpha _i,\delta _i\}$,
$i=1,\dots,M$, and a cosmic-ray experiment.  Depending on the
geographical location, field of view and observation time this
experiment has different exposures to particular regions of the
sky. This can be parametrized by the direction-dependent differential
exposure $dA(\alpha,\delta )$, which we normalize to one,
\[
\int \! dA(\alpha ,\delta )=1.
\]
The integration here runs over the whole celestial sphere.

Under certain assumptions\footnote{Poisson distribution of the number
of observed pairs ``source - cosmic ray'' results from the statistical
independence of such pairs. If there are two close sources in the
catalog, they {\em both} may fall within the angle $\theta$ from the
arrival direction of a cosmic-ray particle. This breaks the above
statistical independence and may cause deviations from the Poisson
distribution. Thus, the Poisson distribution works when the number of
close (as compared to $\theta$) pairs of sources within the acceptance
area is small. One may modify slightly the counting procedure to give
quantity which is distributed exactly according to Poisson
distribution. Namely, one may count {\em cosmic-ray events} which fall
within a given angular distance $\theta$ from any of candidate sources
(so the cosmic-ray events which are close to two candidate sources are
not counted twice). We do not use this version of the procedure here
for the reasons of compatibility with previous results.}, the number
$n$ of observed pairs ``source -- cosmic ray'' separated by angular
distances smaller than some (small) angle $\theta $ is described by
the Poisson distribution,
\begin{equation}
P_{n}\left[S,B\right] =\frac{(S+B)^{n}}{n!} \cdot\e^{-(S+B)}\;,
\label{Poisson}
\end{equation}
where $B$ denotes the average number of background events (that is,
either originated from the sources not contained in the catalog or
deflected) and $S$ denotes the average number of events
which originate from the sources in the catalog (we call $S$ the signal).

The background events are supposed to be isotropically distributed and
detected according to the experiment's exposure, hence
\begin{equation}
\label{Bdef}
B(\theta) = N\sum_i \int_{\Omega_i} dA,
\end{equation}
where the sum is taken over all sources in the catalog, $\Omega_i$ is
a circle of the radius $\theta$ centered on $i$-th source and $N$ is
the total number of cosmic-ray events in the data set.
This can be written as
\begin{equation}
\label{B}
B(\theta) = \frac{\pi \theta^2}{4\pi} \, N M F,
\end{equation}
where 
\begin{equation}
F\equiv \frac{4}{M \theta^2} \sum_i \int_{\Omega_i} dA
\label{eq:F}
\end{equation}
is a geometrical ``correction factor'' which describes the
distribution of the objects from a given catalog, with respect to the
acceptance of the experiment: it equals one for the uniform
distribution and is smaller (larger) than one if there is an
underdensity (overdensity) of sources within the acceptance area. For
small $\theta$ it does not depend on $\theta$.  The exact values of
$F$ for particular experiments and a particular catalog of sources can
be obtained from Eq.~(\ref{eq:F}); for the BL Lac sample under
discussion and some ongoing and planned experiments they are listed in
Table~\ref{tab:exp}.
\begin{table}
\begin{tabular}{|c||c|c|c|}
\hline
Experiment      & $\sigma$  & Ref.                           &$F$  \\
\hline
HiRes (stereo) & $0.6^\circ$ & \cite{HiRes,HiRes:exposure}&1.38\\
 \hline
AGASA          & $2.4^\circ$ & \cite{AGASA}               &1.37\\
PAO (surface)   & $1.4^\circ$ & \cite{PAO1,PAO2}           &0.53\\
PAO (hybrid)  & $0.6^\circ$ &                              &0.48\\
TA (surface)& $1.55^\circ$ & \cite{TA}                    &1.41\\
TA (hybrid)  & $0.62^\circ$ &                              &1.51\\
  \hline
\end{tabular}
\caption{Parameters of the cosmic-ray experiments.
$\sigma$ is the angular
resolution of the experiment (the radius of the circle containing 68\% of
the reconstructed events from a point source).
Reference is given for the value of $\sigma $ and for the details of
direction-dependent exposure.
See Eq.~\eqref{eq:F} for the definition of $F$.
\label{tab:exp}
  }
\end{table}

The signal $S$ is  different for
different experiments. Let us express it in terms of the fraction
$\eta$. For $S\ll N$, the expression is linear\footnote{In what follows,
we will need to integrate over all possible values of $\eta$ up to 1,
where the linear approximation~(\ref{eq:S-of-eta}) fails. In practice,
one may always reexpress the formulae written in terms of $\eta$
through the signal $S$ and integrate over $0\le S \le\infty$.}. The
coefficient of proportionality depends on the acceptance of the experiment,
the angular resolution (including the shape of the point-spread function)
and on the angle at which the signal is measured. The exact expression
reads
\begin{equation}
S(\theta) = g(\theta) F  N \eta,
\label{eq:S-of-eta}
\end{equation}
where the factor $g(\theta)$ is the integrated point-spread function of the
experiment: this is the fraction of the events observed within the
angle $\theta$ from the position of the point source. For instance,
for $\theta$ equal to the angular resolution (by definition, the
radius of a circle containing 68\% of the events) one has
$g(\theta_{68}) = 0.68$~\footnote{Since at $\eta\sim1$ the linear
approximation~(\ref{eq:S-of-eta}) fails, final results depend slightly
on the shape of the point-spread function (in this paper, we assumed
two-dimensional Gaussian distribution.}.

In a given experiment one measures $n$ and has to determine $\eta$.
The likelihood function for $\eta$ can be obtained from
Eq.~(\ref{Poisson}) by normalizing $P_n$ to 1, $\int_0^1{\cal L}(\eta)
d\eta =1$. One has 
\begin{equation}
{\cal L}\left(\eta\right)=
\frac{gFN}{\Gamma(n+1,B)} \l gFN\eta+B \r^n
\e^{-gFN\eta- B}.
\label{likelihood}
\end{equation}
For the original HiRes stereo sample \cite{HiRes}, $B=3.5$ and $n=11$
at $\theta = 0.8^\circ$ \cite{BL:HiRes}. At the 95$\%$~C.L. one has
$0.015 < \eta < 0.035$. Here and in the estimates of the next section
we have set $\theta = \sqrt{2}\sigma$, where $\sigma$ is the angular
resolution of the relevant experiment (for the discussion of the
optimal choice of $\theta$ see Refs.~\cite{BL1,BL:HiRes} and
Sec.~\ref{sec:Predictions}).

%%%%%%%%%%%%%%%%%%%%%%%%%%%%%%%%%%%%%%%%%%%%%%%%%%%%%%%%%%%%%%%%%%%%%%%%

\section{Predictions for future experiments}
\label{sec:Predictions}
With two different experiments X and Y, one obtains two likelihood
functions, ${\cal L}_X(\eta )$ and ${\cal L}_Y(\eta)$. The
two-dimensional probability distribution function corresponding to the
observation of $\eta _1$ in the experiment X and $\eta _2$ in Y is
simply
\[
{\cal L}_X(\eta_1 ){\cal L}_Y(\eta_2).
\]
This formula provides a starting point for quantitative analysis of
the compatibility of the two experiments. Changing variables to
$\eta_1 + \eta_2$ and $\eta_1-\eta_2$ and integrating over $\eta_1 +
\eta_2$ gives the probability distribution for the difference
$\eta_1-\eta_2$ which tells one how well the two experiments are
compatible with each other within our hypothesis. This assumes that
the number $n_Y$ of pairs ``source -- cosmic ray'' separated by the
angle $\theta_Y$ observed in the experiment $Y$ is known.

Alternatively, assuming the likelihood function of $\eta$ as
follows from the experiment $X$, Eq.~(\ref{likelihood}), one may
predict the number of pairs $n_Y$ which should be observed by
the experiment $Y$ if the hypothesis is correct. The probability
distribution for this number  is
\[
P_{n_Y}= \int\limits_0^1\!d\eta \,{\cal L}_X(\eta,n_X )
\frac{( g_YF_YN\eta+B_Y )^{n_Y}}{n_Y!}
\e^{-g_YF_YN\eta- B_Y} .
\]
The 95\% CL interval $\Delta n_{95\%}=(n_1,n_2)$ for the number of
pairs to be observed is defined by the conditions
\begin{align*}
%\label{interval}
P\left[n_1\right]=0.025
\;,~~~P\left[n_2\right]=0.975\;,\\
{\rm where}~~~~P\left[m\right]=\sum_{n_Y=0}^{m}P_{n_Y} 
\end{align*}
(to be more precise, $n_1$ is the largest integer for which
$P\left[n_1\right]\le 0.025$ and $n_2$ is the smallest integer for which
$P\left[n_2\right]\ge 0.975$).

\begin{table}
\begin{tabular}{|c||c|c|c|c||c||c|}
\hline
Experiment      & $\theta=\sqrt{2}\sigma $  & $N$ & $S$ & $B$ &  $\Delta
n_{95\%}$ & $N_{\rm 0}$ \\
\hline
HiRes, original &$0.85^\circ$ &271  & 7.46 & 3.54 &         &      \\
\hline
HiRes (stereo) & $0.85^\circ$& 190  & 5.23 & 2.48 & 1-17    & 271  \\
AGASA          & $3.39^\circ$& 1500 & 43.4 & 310  & 308-417 & 3870 \\
PAO (surface)  & $1.98^\circ$& 500  & 5.50 & 13.5 & 10-31   & 3517 \\
PAO (surface)  & $1.98^\circ$& 8000 & 87.9 & 216  & 239-413 &      \\
PAO (hybrid)   & $0.85^\circ$& 150  & 1.86 & 0.68 & 0-7     & 467  \\
PAO (hybrid)   & $0.85^\circ$& 2000 & 24.8 & 9.10 & 15-66   &      \\
TA (surface)   & $2.19^\circ$& 500  & 15.0 & 44.4 & 42-83   & 1560 \\
TA (surface)   & $2.19^\circ$& 8000 & 239  & 710  & 785-1235&      \\
TA (hybrid)    & $0.88^\circ$& 150  & 4.40 & 2.28 & 1-15    & 277  \\
TA (hybrid)    & $0.88^\circ$& 2000 & 58.6 & 30.3 & 47-161  &      \\
\hline
\end{tabular}
\caption{Predictions (95\% CL) for the number $\Delta
n_{95\%}$  of pairs ``cosmic ray --
BL Lac'' separated by less than $\theta$ to be observed by future
experiments for the catalog of 156 BL Lac's.
The sets of UHECR are constrained by the cuts
on energy ($E>10^{19}$~eV) and on zenith angle ($z<60^\circ$ for the
Pierre Auger Observatory (PAO) and the Telescope Array (TA), $z<45^\circ$
for AGASA). $N$ is the total number of events in a
dataset, $S$ and $B$ are the corresponding numbers of average signal and
background coincidences. $N_0$ is the minimal number of events which the
experiment needs to accumulate in order to reach the current HiRes
sensitivity.
\label{tab:main}
}
\end{table}
In Table~\ref{tab:main} we present the expected number of pairs
``source -- cosmic ray'' for several future cosmic-ray data sets. 
The prediction is done for 
typical total numbers of events $N$ expected to be observed. It is
presented in the form of the 95\% CL interval $\Delta
n_{95\%}=(n_1,n_2)$. If the number of actually observed pairs falls
within the indicated interval, it would be compatible with our
hypothesis, otherwise it would exclude it with the probability 95\%.

The ability to test the hypothesis is affected by many properties of
the experiments. The most important ones are the angular resolution
and the total number of events; however, favorable location in the
Northern hemisphere (with more known BL Lacs in the field of view)
also provides a noticeable advantage. These parameters can be combined
into a single quantity --- signal-to-noise ratio $Q$ --- which
characterizes the ``sensitivity'' of a given experiment to the BL Lac
signal. It is defined as the ratio of the signal $S$ to the typical
fluctuation of the background, 
\begin{equation}
Q = \frac{S}{\sqrt{B}} = \frac{\sqrt{NF}}{\theta} \cdot
\frac{2 g(\theta)\eta}{\sqrt{M}}.
\label{eq:sensitivity}
\end{equation}
The absolute value of $Q$ is related to the significance at which the
existence/absence of the signal can be established. Namely, $Q\gg 1$
is required to test the presence of the signal\footnote{The value of
$Q$ depends strongly on the angle $\theta$. The latter should be
chosen in such a way as to maximize the signal significance. In the
cases considered in Refs.~\cite{BL1,BL:GMF,BL:EGRET,BL:HiRes} this
optimal angle was $\theta\approx\sqrt{2}\sigma$.}. More importantly,
different experiments can be {\em compared} in this way: the one with
larger $Q$ is more sensitive to the signal. Assuming that $\theta$ is
chosen in the same way for all experiments (for instance, $\theta =
\sqrt{2}\sigma$ as in the calculations above), the only
experiment-dependent factor in $Q$ is the first factor in
Eq.~(\ref{eq:sensitivity}). This factor determines the scaling of $Q$
with the total number of events, the optimal angle $\theta$ and the
geometrical correction factor $F$. Thus, Eq.~(\ref{eq:sensitivity})
allows to estimate the number of events which are necessary for the
two experiments to have the same signal-to-noise ratio.

Clearly, in order to test the hypothesis formulated on the basis of
the HiRes data, another experiment has to have at least similar
sensitivity. It is interesting to see how many events the experiments
discussed above have to accumulate to reach the same signal-to-noise
ratio as the HiRes experiment. These numbers are presented in
Table~\ref{tab:main}. One can see, for instance, that the Pierre Auger
experiment will have to collect of order 500 events with
$E>10^{19}$~eV in the hybrid mode (and of order 3500 events observed
by the surface detector only) in order to fully probe the hypothesis
formulated here.

\section{Conclusions}

To summarize, in this paper we have formulated, on the basis of the
HiRes data, the quantitative hypothesis about the connection between
UHECRs and BL Lacs (Sect.~\ref{sec:hypo}, (i)--(iii)) and defined the
procedure to test this hypothesis in future experiments. The results
are presented in Table~\ref{tab:main}. As follows from these results,
the strong correlations observed in the HiRes data are consistent, at
the 95$\%$~C.L., even with the absence of signal in some data sets. In
particular, the final AGASA data set and the first release of the
Pierre Auger data, both expected in near future, may not exhibit any
correlations with this sample of BL Lacs and still be consistent with
the correlation hypothesis formulated in Sec.~\ref{sec:hypo}. On the
contrary, the absence of correlations in the future Telescope 
Array and full Pierre Auger data may allow one to falsify this
hypothesis. This difference is due to different angular resolutions,
exposures and locations of these experiments. Several factors (e.g.,
the choice of the optimal angle $\theta$ with account of the shape of
the individual point-spread function) may improve the sensitivity of
the experiments compared to the indicative estimates given in
Table~\ref{tab:main}.

In principle, other claims of correlations between UHECR and BL Lacs
\cite{BL1,BL:GMF,BL:EGRET} can be tested quantitatively in a similar
way.  However, due to the cut adjustment (compensated by the penalty
factor) to the best signal in these cases the simple analytic relation
which in fact determines the flux of cosmic rays produced by BL Lacs,
Eq.~(\ref{likelihood}), is no longer valid; its analog may be
calculated numerically. We leave this question for future work.  

The correlations of BL Lacs with the HiRes stereo data, if confirmed,
would imply that a few-percent fraction of the highest-energy
cosmic-ray flux is composed of neutral particles coming from
cosmological distances. This new UHECR puzzle would be as difficult to
solve within conventional physics as the celebrated GZK
cutoff~\cite{GZK} problem.

We are indebted to K.~Belov, M.~Giller and D.~Semikoz for helpful and
stimulating discussions. This work was supported in part by the INTAS
grant 03-51-5112 (D.G.,  P.T.\  and S.T.), by the grant of the President
of the Russian Federation NS-2184.2003.2, by the grants of the Russian
Science Support Foundation (D.G.\ and S.T.) and by the fellowships of the
"Dynasty" foundation (awarded by the Scientific board of ICFPM; D.G.\ and
S.T.). The work of P.T. is supported in part  by IISN, Belgian Science
Policy (under contract IAP V/27).

\end{document}